\documentclass[letterpaper]{article} % DO NOT CHANGE THIS
\usepackage{aaai24}  % DO NOT CHANGE THIS

\usepackage{times}  % DO NOT CHANGE THIS
\usepackage{helvet}  % DO NOT CHANGE THIS
\usepackage{courier}  % DO NOT CHANGE THIS
\usepackage[hyphens]{url}  % DO NOT CHANGE THIS
\usepackage{graphicx} % DO NOT CHANGE THIS
\usepackage{natbib}  % DO NOT CHANGE THIS AND DO NOT ADD ANY OPTIONS TO IT
\usepackage{caption} % DO NOT CHANGE THIS AND DO NOT ADD ANY OPTIONS TO IT
\usepackage{algorithm}
\usepackage{algorithmic}

\usepackage{newfloat}
\usepackage{listings}
\DeclareCaptionStyle{ruled}{labelfont=normalfont,labelsep=colon,strut=off} % DO NOT CHANGE THIS
\floatstyle{ruled}
\newfloat{listing}{tb}{lst}{}
\floatname{listing}{Listing}
\title{Interpretable Classification of Early Stage Parkinson's Disease from EEG}

\author{
   Amarpal Sahota\textsuperscript{\rm 1},  Amber Roguski\textsuperscript{\rm 1},
    Matthew W Jones\textsuperscript{\rm 1},\\
    Michal Rolinski\textsuperscript{\rm 2},
    Alan Whone\textsuperscript{\rm 1},
    Raul Santos-Rodriguez\textsuperscript{\rm 1},
    Zahraa S. Abdallah\textsuperscript{\rm 1}
}
\affiliations{
    \textsuperscript{\rm 1}University of Bristol, UK\\
     \textsuperscript{\rm 2}Torbay Hospital, UK\\
    amarpal.sahota@bristol.ac.uk, zahraa.abdallah@bristol.ac.uk, enrsr@bristol.ac.uk

}

\begin{document}

\maketitle

\begin{abstract}

Detecting Parkinson's Disease in its early stages using EEG data presents a significant challenge. This paper introduces a novel approach, representing EEG data as a 15-variate series of bandpower and peak frequency values/coefficients. The hypothesis is that this representation captures essential information from the noisy EEG signal, improving disease detection. Statistical features extracted from this representation are utilised as input for interpretable machine learning models, specifically Decision Tree and AdaBoost classifiers. Our classification pipeline is deployed within our proposed framework which enables high-importance data types and brain regions for classification to be identified. Interestingly, our analysis reveals that while there is no significant regional importance, the N1 sleep data type exhibits statistically significant predictive power ($p < 0.01$) for early-stage Parkinson's Disease classification. AdaBoost classifiers trained on the N1 data type consistently outperform baseline models, achieving over 80\% accuracy and recall. 
Our classification pipeline statistically significantly outperforms baseline models indicating that the model has acquired useful information.  Paired with the interpretability (ability to view feature importance's) of our pipeline this enables us to generate meaningful insights into the classification of early stage Parkinson's with our N1 models. In Future, these models could be deployed in the real world - the results presented in this paper indicate that more than 3 in 4 early-stage Parkinson's cases would be captured with our pipeline. 

\end{abstract}

\section{Introduction}
Parkinson's Disease (PD) is a neurodegenerative disease characterised by degeneration of the dopaminergic neurons in the substantia nigra pars compacta of the brain \cite{yuvaraj2018novel}. It is the second most prevalent neurological disorder after Alzheimer's disease \cite{han2013investigation}. Symptoms of Parkinson's include muscle stiffness, slowness of movement (bradykinesia), physical instability and dysphonia (voice disorder) \cite{yuvaraj2018novel}. Timely diagnosis of the disease is challenging as symptom patterns are inconsistent across the disease population with doctors often diagnosing patients after the onset of motor symptoms. By this stage, individuals have lost up to 60\% of the dopaminergic neurons in the substantia nigra. Early diagnosis is key to obtaining the best treatment plan for patients \cite{khoshnevis2021classification}. Furthermore, if an early diagnosis could be achieved this would support the development of treatments targeted at supporting the surviving dopaminergic neurons and slowing the development of Parkinson's Disease \cite{pagan2012improving}. 

The severity of Parkinson's disease can be described by a clinical rating scale such as the Hoehn and Yahr (HY) scale  \cite{bhidayasiri2012parkinson}. The HY scale defines broad categories of motor function in the disease and has 5 stages with stage 1 being the earliest. In stage 1 the motor impairment symptoms are mild. Therefore, individuals may not seek medical attention at this stage and if they do doctors are often unable to make the correct diagnosis.

To help achieve a timely diagnosis of Parkinson's, work has been done on utilising Electroencephalography (EEG) data to classify the disease. EEG is a non-invasive, relatively inexpensive method which records the electrical activity of pyramidal cortical neurons in the brain. Furthermore, EEG has a high temporal resolution providing insight into functional processes in the brain. Methods for classification with EEG data in the literature range from feature extraction (Band power, entropy, spectral features) followed by a machine learning model such as a support vector machine, random forest or artificial neural network to the direct application of a deep learning model such as a convolutional neural network. See section \ref{related work} for a detailed review. 

Classifier performance in the literature varies from approximately 70\% to some approaches achieving 90\% +  and even close to 100\% accuracy. However, limited exploration has been done on the earliest stage of Parkinson's Disease or on explainability. A desirable classifier would have high accuracy and recall for early-stage Parkinson's Disease (HY stage I). This would enable real-world deployment to aid diagnosis. Secondly, a desirable classifier would be explainable - identifying why individuals are classified as healthy or PD. This would enable a better understanding of the disease state brain and individual classifications. Finally, it is noted that limited exploration has been done on different types of EEG (e.g. sleep vs wakeful) for the disease, in the literature wakeful EEG data is widely used. Using different types of EEG could improve classifier performance and aid in better understanding the diseased brain.

Our contributions using a novel dataset comprising of EEG data recorded from early stage Parkinson's patients (see \ref{Data} for further details) are twofold. (1) An interpretable pipeline for EEG classification that achieves high performance (Over 80\% accuracy and recall on early-stage PD) and outperforms baseline models. A key component of this pipeline is a novel representation of EEG data as a 15-variate series of band power coefficients and peak frequency values/coefficients as a means of capturing useful information from the noisy raw EEG signal. (2) A framework to determine the importance of individual brain regions and EEG data type (wakeful vs. different stages of sleep) in classification. 

% \vspace{0.3cm}
\section{Related Work} \label{related work}
As existing works, we rely on band power values and peak frequency values from EEG. These are both derived from the power spectral density which represents the strength of a signal across frequencies. Band power values refer to specified ranges within the power spectral density whereas peak frequency refers to locations of the power spectral density where the frequency coefficients are maximal. In \cite{chaturvedi2017quantitative} the authors use EEG band power values in combination with median and peak frequency to achieve 78\% accuracy on PD classification with a balanced data set (50 PD, 40 Healthy Control).  In \cite{betrouni2019electroencephalography}, with similar features, they achieve a classification accuracy of 86\% with a support vector machine. Band power and peak frequency features have the limitation of discarding phase information of the frequency components as well as reducing the temporal resolution to the window over which the values were calculated. Further, in \cite{waninger2020neurophysiological}  it is proposed to combine coherence statistics (a measure of functional connectivity between regions) with wavelet coefficients and apply linear discriminant function analysis, achieving a classification accuracy of 94\%. More complex spectral features have also been used with promising results in EEG classification for PD. In \cite{yuvaraj2018novel}, the authors extend their method to include features from the third-order spectra of EEG data (bispectrum) achieving an accuracy of 99\% with a support vector machine.  One limitation of the study is that according to the Hans and Yoer Parkinson's Disease severity score, many patients were in stage II and III (not early stage). Recent work \cite{khoshnevis2021classification} has followed this but achieved lower accuracy scores using complex spectral features (85\% with decision tree ensemble model).
Deep learning approaches such as convolutional and recurrent neural networks have also shown high performance (though with limited interpretability) such as in \cite{lee2019deep} where the authors achieve 96.9\% accuracy on PD classification. In summary, early detection of Parkinson's Disease via EEG shows promise. However, in the literature, there does not exist a method that achieves high-performance classification of earliest-stage PD while existing classifiers also have limited explainability.

\section{Background} \label{background}

\subsection{EEG Data Types}

When we refer to EEG data type in this paper - this means the state of the participant during EEG recording. Therefore eyes-closed wakeful EEG recording is one EEG data type. We explore five different EEG data types in this paper, in addition to the eyes-closed wakeful EEG we also assess EEG data from the four sleep stages; N1, N2, N3 and REM. Sleep can be divided into rapid eye movement (REM) and non-rapid eye movement sleep (stages N1-N3).   Each sleep stage (N1, N2, N3, REM) refers to progressively deeper sleep where in a typical healthy night of sleep an individual will cycle through these stages 4-5 times  \cite{sleeppatel2022physiology}. Each sleep stage is different in function and is associated with different brain wave patterns. Thus it is advantageous to study EEG data from all four sleep stages in addition to wakeful data. This approach goes beyond much of the current research literature that focuses on EEG from a single state (commonly wakeful data).  

\subsection{EEG Band Power Calculation} \label{EEG BPW Calculation} 
EEG data measures brain activity via electrodes placed on the scalp. Research has shown the strength of EEG signals in frequency bands to be indicative of brain activity \cite{bandpowerpaper}. Exact band power definitions vary slightly across research literature, the definitions we use to set the power bands as delta (0.5--4Hz), theta (4--8Hz), alpha (8--12Hz), sigma (12--16Hz), beta (16--30Hz) and gamma (30--40Hz). 

Welch’s method \cite{welch1967use} enables one to calculate the power spectral density (PSD) for a signal in the time domain. This method splits the time series into windows, calculating the periodogram for each window via the Discrete Fourier Transform (DFT) and then averaging across windows. The power spectral densities within the defined power bands are calculated by integrating the area for the corresponding frequency band under the PSD curve. 

Absolute band power refers to the calculation of the power bands as described above. Relative band power normalises the area under the PSD curve across all signal samples to account for signal power differences. Per frequency band, this is calculated by dividing the absolute band power by the total power of the signal.

\subsection{EEG Peak Frequency Values and Coefficients Calculation} \label{Peak Freq Calc} 
The power spectral density (PSD) for the EEG signal is calculated via Welch's method using the Discrete Fourier Transform (DFT) and windowing as described in \ref{EEG BPW Calculation}. The resulting PSD has frequency values and corresponding coefficients referring to the strength of those frequencies. The PSD is normalised by dividing all coefficients by the sum of coefficient values. This ensures that the PSD coefficients are comparable across EEG signals of differing amplitudes.

When analysing the PSD for peak frequency we ignore values below 1Hz as is typical for EEG analysis. The four frequency values with the highest coefficients and their respective coefficient values are the resultant features. These features give fine-grained information on power distribution across peak frequency values in the PSD.

\section{Methods} \label{methods}

The data used for this work was from an unpublished dataset
% \footnote{Process to access the data will be available in the published version of this work. See proceedings of the 8th International Workshop on Health Intelligence (W3PHIAI-24)} 
containing 57 channel EEG data of five different types (Wakeful, N1, N2, N3, REM) for Parkinson's Disease and Healthy Control participants (see section \ref{Data} for further details on the data).

\subsection{Preprocessing}
EEG data were manually scored in 30-second epochs according to AASM scoring guidelines \cite{berry2017aasm}, thus labelling data segments as Wake, NREM stages 1-3, REM or Artefact. EEG data were down-sampled from 512Hz to 256Hz and bandpass filtered between 0.25 – 40Hz. Further artefact removal and removal of bad channels were done via Independent Component Analysis and manual inspection of the signal and power spectral density plots. After excluding artefact components from the data, bad channels were interpolated. The data was then re-referenced using the REST referencing technique \cite{dong2017matlab,yao2001method}. The resultant pre-processed data set therefore contained artefact-free Wakeful, N1, N2, N3 and REM labelled  57 channel EEG data across all participants. 

\subsection{A Time Series Pipeline for EEG Classification }\label{classifier}
A key component of our classification approach is to use a novel representation of the EEG signal. The representation is achieved by transforming a single EEG signal into a 15-variate series of features per window (we select a window size of 6 seconds). The 15-variate series are defined as relative power across frequency bands (7 features) as well as the four peak frequency coefficients and the corresponding four peak frequency values. We hypothesise that this feature series summarises useful information from the noisy raw EEG signal. Calculations for these features are outlined in \ref{EEG BPW Calculation} and \ref{Peak Freq Calc}. 

Traditional approaches would simply calculate the value of these features for the entire EEG signal. We calculate the statistics of these 15 features that describe how they change over time throughout the duration of the EEG signal recording. We hypothesise that this added information will increase classifier performance.

To calculate the statistics of the feature series we use the library TS-Fresh \cite{christ2018time}. There are two settings we use. The first is 'Minimal' which calculates 15 features from the input series including mean, variance, sum, maximum, minimum and others. The second setting 'Efficient' calculates up to 700 features per series with the features describing the series in more detail including auto-correlation of the series across different lags, entropy, and features describing the absolute value of consecutive changes in the variable.

For final classification, the statistical features are input to classification models (Decision Tree and Ada Boost). These models are advantageous in being interpretable which is very important in healthcare settings.

\begin{figure}[h!]
        \centering
        \includegraphics[width=0.95\columnwidth]{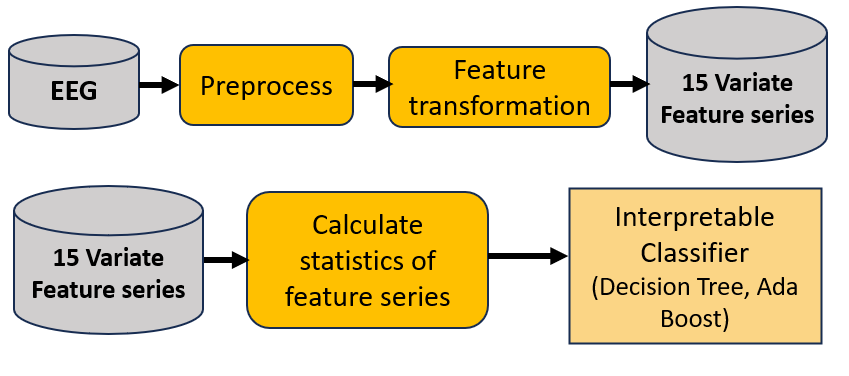}
        \caption{Classification pipeline. EEG data is preprocessed and transformed into a 15-variate series of bandpower coefficients, peak frequency coefficients and peak frequency values. Statistics of these series are used as input into an interpretable classifier (AdaBoost / Decision Tree).}
        \label{classifier_fig}
\end{figure}

\subsection{A Framework for Understanding the Importance of EEG Data Types and Brain Regions} \label{framework}
In order to offer not only a prediction on an individual example but also informative insights, we aim to identify the key EEG data types and brain regions that are indicative of the presence of the disease. 

\begin{figure}[h!]
        \centering
        \includegraphics[width=0.9\columnwidth]{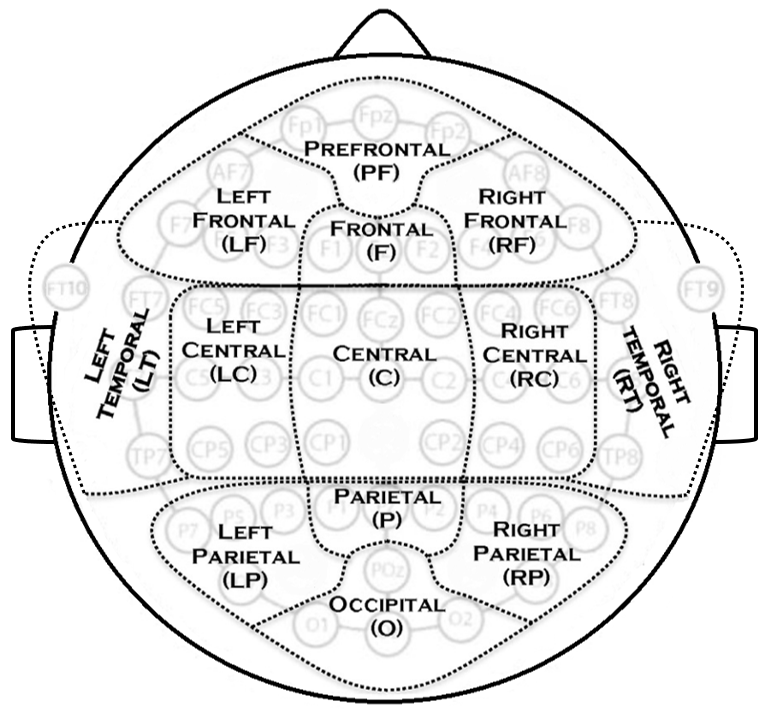}
        \caption{EEG channels grouped into 13 brain regions of interest, figure adapted from \cite{regionaleeg}.}
        
        \label{brain_regions_fig}
\end{figure}

We assume that the EEG data is tagged by data type - in our study, we consider five (wakeful, N1, N2, N3 and REM). Secondly, the 57 EEG channels are grouped into 13 brain regions of interest as shown in Figure \ref{brain_regions_fig}. We train a classifier in each region for each data type – resulting in 65 results (13 regions x 5 data types) per classifier. Analysing these results per region per data type enables the practitioner to determine which brain regions and which data types are more influential on the predictions. We assume in what follows that a higher performance (accuracy, recall, precision) implies higher relevance for that brain region and EEG data type for discriminating between classes.

Addressing brain regions and EEG data types in this way is expected to lead to a better disease state understanding by revealing the importance of individual brain regions across EEG data types for classification as the disease progresses. Furthermore, this framework could also enable more efficient implementation of classifiers and labelling activities in practice. For example, if a known region during wakefulness is deemed sufficiently discriminatory for a disease, then in practice one could collect EEG data from this region only during wakefulness to diagnose patients.

\section{Experiments and Discussion} \label{results}

\subsection{Data \label{Data}}

The dataset records 57-channel EEG data during wakefulness and the four stages of sleep (N1, N2, N3, REM) for the study participants. Study participants consist of Healthy Control (HC)  and Parkinson's Disease (PD) participants as well as individuals with REM Sleep Behaviour Disorder (RBD) and individuals with both RBD and PD. This paper focuses on data from HC participants and participants with PD only.

The local Parkinson’s UK Patient and Public Involvement (PPI) group were consulted on the proposed study protocol and all study procedures were completed in accordance with the Declaration of Helsinki.  In home Electroencephalography (EEG) data recording was conducted on multiple (2 or 3) nights for all participants in the study. Electrophysiological data was recorded from 57 scalp locations defined by the 10-20 system \cite{klem1999ten} and in line with AASM standards for polysomnography \cite{berry2017aasm}. The demographics of Healthy Control and Parkinson's Disease study participants do not vary significantly - see Table \ref{data_stats}. Furthermore, socioeconomic status indicators (accommodation status, vehicle access, employment status \& position) and health status predictors (years of education, marital status) were also closely matched between groups.

\begin{table}[h!]

\centering
\begin{tabular}{ |c|c|c|c| } 
  \hline
 - & Healthy Control (19) &  Parkinson's (11)  \\
 \hline
 Female & 5 (26.3\%) &  4 (36.4\%)   \\
 Male & 14 (73.7\%) &  7 (63.6\%)  \\
 Age & 69.57 $\pm$ 8.77 & 67.82 $\pm$ 10.77  \\ 
 HY Stage & - &  1.55 $\pm$ 0.21  \\ 

\hline
\end{tabular}
\caption{Demographics of study participants. \\
}
\label{data_stats}
\end{table}
As described in Table \ref{data_stats}, EEG data was recorded on multiple nights. Therefore in cases where usable EEG data was recorded on multiple nights the number of samples exceeds the number of participants. Pre-processed EEG data statistics are shown in Table \ref{data_stats_2}.

\begin{table}[h!]
\begin{center}
\begin{tabular}{ |c|c|c| } 
  \hline
 EEG Type  &  Samples & Length (epochs)  \\
 \hline
 Wakeful  &  HC 17 , PD 11 & 10 $\pm$ 2  \\
 N1   &  HC 26 , PD 15 & 30 $\pm$ 23 \\
 N2  &  HC 26 , PD 15 & 300 $\pm$ 110\\ 
 N3  &  HC 26 , PD 15 & 150 $\pm$ 54\\ 
 REM   &  HC 26 , PD 14 & 105 $\pm$ 40 \\ 
\hline
\end{tabular}
\caption{ Statistics on data recorded per EEG data type. The number of samples is greater than the number of participants when one or more participants have EEG data of that type recorded on two different occasions. Demographics per data type will therefore approximately but not exactly equate to the values in Table \ref{data_stats}.}
\label{data_stats_2}
\end{center}
\end{table}

\subsection{Experimental Procedure}

Our classification pipeline (section \ref{classifier}) was implemented within our classification framework (section \ref{framework}). As described in section \ref{classifier} this pipeline first transforms the EEG signal into a 15-variate feature series, then calculates the statistics of those series and finally uses the statistics of those series as input to an Ada Boost Classifier and a Decision Tree Classifier. Two settings were used for the statistics calculated from the feature series - 'Minimal' (~15 features per series) and 'Efficient' (~700 features per series). Further details on all statistics for both the 'Minimal' and 'Efficient' settings can be found in the appendix. The statistics were calculated using the time series classification package TSFresh \cite{christ2018time}. 

Five-fold cross-validation was performed to assess classifier performance. The training set was used to perform a Grid search on hyper-parameters. Hyper-parameter grids for the Decision Tree and Ada Boost spanned key parameters and were set as follows. Decision Tree Grid - 'min\_ samples\_leaf': [1, 2, 3, 5,10], 'max\_depth': [1, 2, 3, 5, None], 'criterion': ["gini", "entropy"], max\_features': [None, 'sqrt']. Ada Boost Grid -  'n\_estimators': [2, 3, 5, 10, 20, 40, 50, 100], 'learning\_rate': [0.01,0.05,  0.1, 0.2, 0.4, 1.0, 2.0, 10.0].

Per our framework, regional classifiers were trained per EEG data type (Wakeful, N1, N2, N3, REM). Baseline models were also deployed. Baseline models used the mean values of the 15-variate series. Namely band-power coefficients for the signal, peak frequency values and peak frequency coefficients. These types of features are typically used in this classification task (Static features with no time-varying information component) \cite{betrouni2019electroencephalography}. Baseline Decision Tree and Ada Boost models were trained with the same aforementioned parameter grids for hyperparameter optimisation. 

Finally, dummy models with no knowledge were also deployed with three different strategies ('prior', 'stratified' and 'uniform'). The purpose of the dummy models is to statistically prove that we have learned something in the challenging regime of a few samples, early-stage PD participants and typically noisy EEG data. The purpose of the baseline models is to demonstrate that our approach to statistical time series features of the feature series representation outperforms standard baseline methods.

\subsection{Results} \label{Results}
\begin{figure*}[h!]
        \centering
        \includegraphics[width=0.8\textwidth]{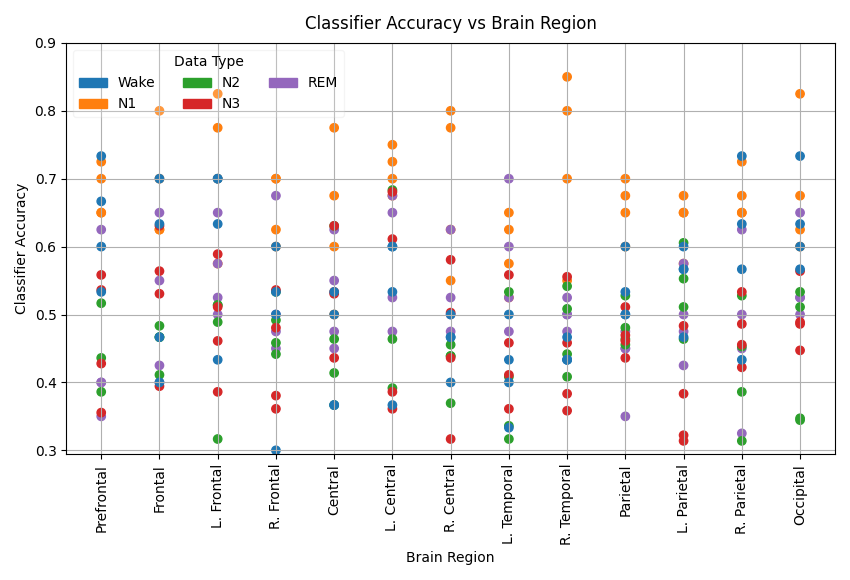}
        \caption{Classifier accuracy vs Brain region with data points coloured by their data type. Visibly there are no clear regional trends but in terms of data type, N1 (orange) has the highest performance.}
        
        \label{accuracy_graph}
\end{figure*}

Across EEG data types our pipeline showed the best performance on N1 data versus other data types. However, there was no significant performance difference across regions. Classifier accuracy results per brain region are coloured by data type in Figure \ref{accuracy_graph}. Note that there are four classifiers deployed per region per data type - an Ada Boost model and a Decision Tree that use 'Efficient' or 'Minimal' statistical features. 

In addition to accuracy, we consider recall and precision to be important performance metrics. Recall is a very important metric as it is important to correctly capture early-stage PD even if there are a few false positives. These false positives could then be resolved through further diagnostic tests. 

\paragraph{Best Performing Classifiers}
Considering all performance metrics the best-performing models used the efficient statistical time series features in N1 EEG data from the Left Frontal and Right Temporal brain regions as input to Ada Boost models. The classifier trained on the Right Temporal brain region achieved 85\% accuracy, 85\% precision and 73\% recall. The classifier trained on the Left Frontal brain region achieved a performance of 78\% accuracy, 67\% precision and 80\% recall. 

Baseline models performed considerably worse than our proposed classification pipeline. The highest accuracy of baseline models achieved 73\% accuracy, 70 \% precision and 58\% recall. The highest recall baseline model achieved 60\% accuracy, 50\% precision and 67\% recall.  This outperformance of our pipeline versus baseline models demonstrates the benefits of using statistical features of the transformed 15-variate series as opposed to using features directly from the raw EEG data. 

The best accuracy of dummy models was the 'prior' model which simply predicts the most common class, always. This achieved an accuracy of 62.5\% with a recall of 0\% (precision undefined in the case of 0 predicted positives). The highest recall dummy model was the 'stratified' model which predicts sample classes randomly according to the distribution of labels seen by the model in the training set. This model was run 65 times for fair comparison vs our pipeline and baselines to represent the 13 brain regions across 5 data types. The best-performing dummy 'stratified' model achieved an accuracy of 53\%, a precision of 42\% and a recall of 42\%. 

\paragraph{Statistical Significance of the Results}
Statistical significance tests were conducted with the Friedman and Nemenyi tests using a p-value of 0.05 \cite{pereira2015overview}. Using classifier accuracy, precision and recall as inputs to the test the N1 data type was found to be statistically significantly better than all other data types for accuracy and precision. N1 data was also found to be statistically significantly better than N2, N3 and REM data in recall.  Wake data was statistically significantly better than N2 and N3 data in the recall performance metric. There was no notable statistical significance across brain regions.

N1 data as the most important data type for early stage PD classification is an interesting finding. N1 is the lightest sleep stage and involves a transition from wakefulness to deeper sleep stages. N1 data as the best data type for classification could indicate that early-stage PD participants have a disturbed N1 sleep state with potential small intrusions of wakefulness. Research by \cite{zhang2021nrem}
and \cite{memon2023quantitative} has demonstrated changes in the properties of both NREM and REM EEG for Parkinson's diseased individuals. The statistical outperformance of N1 data that we find should be further explored in future studies.

\paragraph{Analysis of Best Classifiers}
As previously stated the two overall best classifiers used efficient statistical features from N1 data as input to Ada Boost models. They were trained on the Right Temporal and Left Frontal brain region(s). 

The confusion matrix for the Right Temporal classifier is shown in figure \ref{CM} below. This classifier achieves 85\% accuracy, 85\% precision and 73\% recall which is remarkable especially when participants are at an early stage of Parkinson's. Considering the classifier recall, around 3 in 4 early-stage PD cases would be correctly identified. False positives are also minimal with 85\% precision. A refined version of this model could therefore have an impact in practice in identifying early-stage PD and helping to address the diagnostic issue with the disease. 

The Left Frontal classifier has worse accuracy and precision but higher recall at 80\%. Such a classifier could also be deployed in practice but one would have to consider the cost of false positives before deployment. 

\begin{figure}[h!]
        \centering
        \includegraphics[width=0.9\columnwidth]{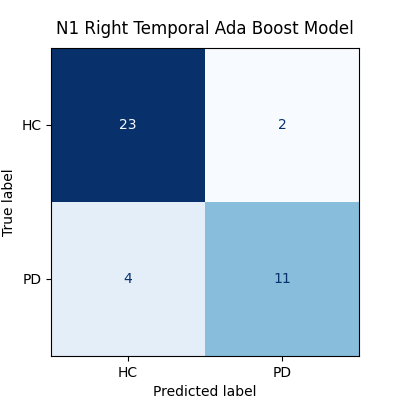}
        \caption{Confusion Matrix for highest accuracy model (85\%). This model uses the 'Efficient' statistical features from the Right Temporal Brain region on N1 data, input to an Ada Boost Classifier.}
        
        \label{CM}
\end{figure}

\begin{figure*}[h!]
        \centering
        \includegraphics[width=1\textwidth]{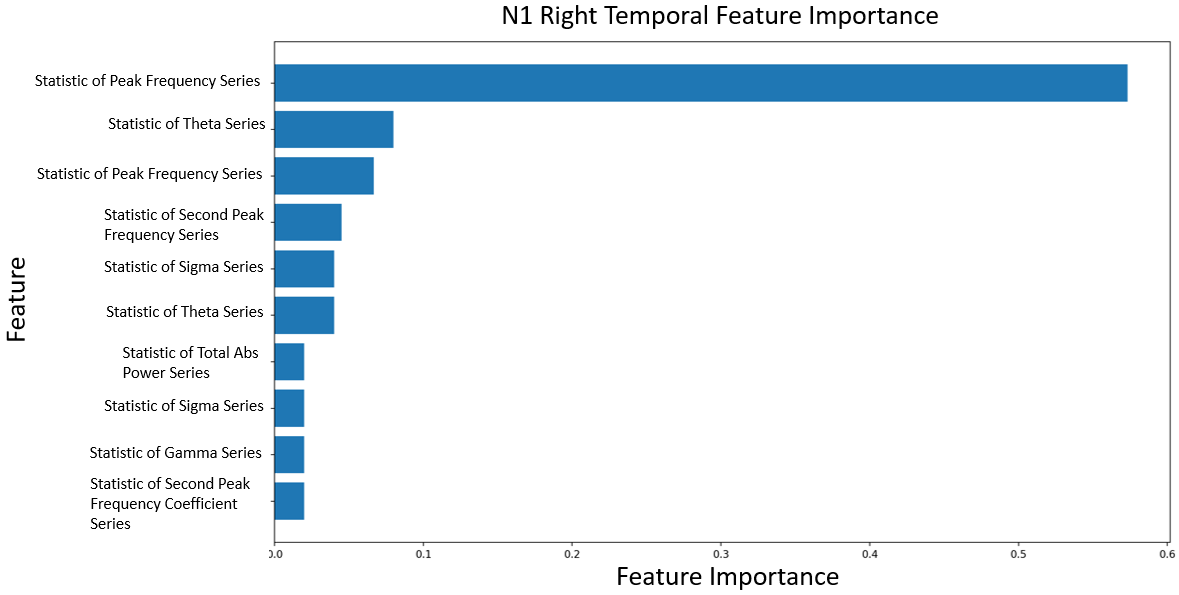}
        \caption{Feature Importance graph, Series refers to which of the 15-variate series the feature was calculated from. Specific statistics corresponding to the top 5 most important features are detailed in the appendix. \ref{Appendix} }
        
        \label{feat_imp}
\end{figure*}

In addition to performance, our model is interpretable. We plot the Ada Boost model feature importance for the Right Temporal N1 Ada Boost Model in Figure \ref{feat_imp}. 

The feature importance graph shows the 10 most important features of the Ada Boost model. Interpretability is particularly important in the healthcare context. The graph shows the feature series that the statistic was calculated from as well as the specific statistic. Notably, the specific statistics are not easily interpretable. This should be addressed in Future Work and is discussed in further detail in section \ref{Future Work}. However, it is notable to see which series the most important features are derived from. 

The most important feature in this model for classification is derived from the Peak Frequency series. Peak frequency has been shown in many EEG classification problems to be an important feature for classification and differences in peak frequency between healthy controls and patients with cognitive impairment have previously been observed.

The subsequent most important features are derived from the Theta, Peak Frequency, Second Peak Frequency, Sigma and Theta feature series respectively. 85\% of the feature importance mass is contained within the top 6 features. These subsequent features re-iterate the importance of the peak frequency series in classification but also highlight the importance of the Theta and Sigma feature series. This aligns with studies that show a shift in the frequency distribution of EEG during sleep for Parkinson's Disease \cite{memon2023quantitative}. 

Thus, in summary, the model we have presented would hypothetically be great in aiding the diagnosis of early-stage Parkinson's Disease. Model performance indicates that 3 in 4 early-stage PD cases would be captured. In addition, the model interpretability as demonstrated through the feature importance discussion would be an important aid to healthcare practitioners.

\subsection{Future Work} \label{Future Work}
There are many clear directions for future work. Firstly, model interpretability can be extended by ensuring that the features used are easier to interpret. As a next step, we hope to improve the interpretability of the features themselves whilst maintaining model performance. This could involve additional feature engineering based on neuroscience literature. For example, sleep spindle density would be a good additional feature. Sleep spindles are characteristic oscillations that occur usually during N2 sleep. Researchers in \cite{memon2023quantitative} for example used an algorithm to analyse sleep spindle density as well as sleep spindle amplitude and other properties. Fewer statistical features and more interpretable curated features could lead to better results as well as better explainability.

Secondly, global brain features can be used to optimise model performance. In this paper, we have used regional features. However, in future, we could use features from across brain regions to improve overall model performance. We can also combine statistical EEG features with brain connectivity metrics that characterise interactions between brain regions. Literature has shown metrics such as Phase Lag Index, Mutual Information, Coherence and others to be good at characterising brain connectivity \cite{chiarion2023connectivity}. These features would also naturally be interpretable with feature importance plots able to highlight clearly which electrode-electrode interactions were most informative for classification. 

Finally, we would also deploy our classification pipeline on other EEG data sets. This would improve confidence in our model performance and findings in terms of feature importance for Parkinson's vs Healthy Control classification. Deployment of additional datasets and proof of generalisation of our pipeline to other datasets would be a necessary step ahead of future potential deployment in the real world. 

\section{Conclusion}
We present an interpretable pipeline for EEG classification that uses statistics of a novel 15-variate feature series representation of the original EEG signal. Deployed on a challenging dataset with early-stage PD participants and limited samples we achieved over 80\% accuracy and 80\% recall with our classification approach. Our pipeline outperformed baseline methods by a large margin which achieved best performances of 73\% accuracy with 58\% recall and 60\% accuracy with 67\% recall. Baseline model performance re-iterates the challenging nature of this classification task.

We tested brain region significance and EEG data type significance in classification. We found N1 sleep data to be statistically significantly the most important data type for early-stage Parkinson's classification. This is supported by studies that indicate differences in Parkinson's EEG characteristics during sleep versus Healthy individuals \cite{memon2023quantitative}. The outperformance of N1 sleep in particular should be explored further. There was no brain region significance found in the classification. 

Feature importance analysis of a model with 85\% accuracy, 85\% precision and 73\% recall showed statistical features from Peak Frequency, Theta, Second Peak Frequency and Sigma feature series to be most important in classification. 

Future work should extend this modelling to additional EEG data sets to further prove the performance of this classification pipeline and confirm the importance of data types, brain regions and features for classification. Future work will also consider data from the entire brain including connectivity metrics for classification as outlined in \ref{Future Work}. 

\appendix{}
\section{Appendix} \label{Appendix}

\subsection{Statistics of Feature Series} \label{representation}
The Python package TSFresh was used to calculate the statistical features for input to our models. Here, in table \ref{Table_TS_stats} we list the explicit definitions of the 5 most important features shown in figure \ref{feat_imp}

The complete list of 'Efficient' and 'Minimal' statistics can be found by referring directly to the TSFresh package documentation \cite{christ2018time}. 

\begin{table*}[h!]
\begin{center}
\begin{tabular}{ |c|p{4cm}|p{6cm}| } 
  \hline
 Feature &  Statistical Feature Name  & Meaning \\ 
 
 \hline
 1. Statistic of Peak Frequency Series
  & Aggregate linear trend with chunk length 10 (minimum) & Calculates a linear least-squares regression for values of the time series that were aggregated over chunks versus the sequence from 0 up to the number of chunks minus one.

  \\
  \hline
 2. Statistic of Theta Series 
   &  Aggregate linear trend with chunk length 5 (mean) & Calculates a linear least-squares regression for values of the time series that were aggregated over chunks versus the sequence from 0 up to the number of chunks minus one.
 \\
  \hline
 3. Statistic of Peak Frequency Series
  &  Aggregate linear trend with chunk length 10 (minimum) & Calculates a linear least-squares regression for values of the time series that were aggregated over chunks versus the sequence from 0 up to the number of chunks minus one.
  \\ 
  \hline
 4. Statistic of Second Peak Frequency Series
  & Quantile q\_0.4 & Calculates the value of the series that is greater than 40\% of the ordered values of the series
\\ 
  \hline
 5. Statistic of Sigma Series & Aggregate linear trend with chunk length 5 (minimum) & Calculates a linear least-squares regression for values of the time series that were aggregated over chunks versus the sequence from 0 up to the number of chunks minus one.
\\ 
\hline
\end{tabular}
\caption{ Further detail on the top 5 most important statistics per the Left Frontal N1 Ada Boost model. The exact calculations for these statistics are available in the TSFresh library documentation by referring to the statistical feature names provided. \cite{christ2018time}. 
}
\label{Table_TS_stats}
\end{center}
\end{table*}

\vspace{5cm}
\bibliography{aaai24}

\end{document}